\documentclass{article}
\usepackage{frascatiphys,here,graphicx,subfigure}
\begin{document}
\title{EXOTICS MESONS: STATUS AND FUTURE
}
\author{
Eberhard Klempt \\{\em Helmholtz-Institut f\"ur Strahlen- und
Kernphysik Nu\ss allee 14-16 D53115 Bonn } }\maketitle
\baselineskip=11.6pt

\begin{abstract}
The evidence for the existence of mesons with exotic quantum numbers
and of hybrid candidates with non-exotic quantum numbers is
critically reviewed, including candidates with hidden charm. Aims
and methods of future searches for hybrid mesons are briefly
discussed.
\end{abstract}
\baselineskip=14pt
\section{Introduction}
The search for exotic mesons is at a turning point. The experiments
at BNL, Protvino, and at LEAR which have reported evidence for
exotic mesons have terminated data taking; data analysis is
completed and the results are published since a few years. On the
other hand, new experiments are ahead of us, COMPASS at CERN and
BESIII in the immediate future, the Hall-D experiment at the
upgraded Jlab facility and PANDA at GSI in the medium-range future.
Hence it seems timely to review the status of exotic mesons to
define the platform from which the new experiments are starting. It
is custom to start from the assumption that glueballs and hybrids
are firmly predicted by Quantum Chromo Dynamics, and experimental
results have to concur with this prediction. Here, a different view
will is adopted: the question is asked if a convincing argument can
be made that the existence of exotic mesons, of hybrids and/or
tetraquark mesons, can be deduced unambiguously from past
experiments. The search for hybrids is part of the wider quest to
understand the role of gluons in spectroscopy\cite{Klempt:2007cp}.

\section{Exotic mesons}
\subsection{Flavor exotic mesons}
Flavor exotic states have a flavor configuration with a minimum of
four quarks like doubly charged states ($uu\bar s\bar d$) or
tetraquark states with heavy flavor ($cs\bar u\bar d$). By
definition, such states cannot mix with regular $q\bar q$ states. In
light-quark meson spectroscopy, there is no accepted flavor exotic
candidate (see also\cite{Filippi:2001gs}). A $c\bar cu\bar d$
candidate will be discussed below.
\subsection{Spin-parity exotics}

Spin-parity exotic mesons have quantum numbers $J^{PC}$ which are
not allowed for fermion-antifermion systems,  $J^{PC}_{exotics}=
0^{--},~~ 0^{+-},~~1^{-+},~~2^{+-},~~3^{-+} \cdots$~. The quantum
numbers $1^{-+}$ are part of the series $0^{-+}$, $1^{-+}$,
$2^{-+}$, $\cdots$~; the isovector states are called $\pi$, $\pi_1$,
$\pi_2$, $\pi_3$ $\cdots$~. In this series, the $J$-odd states are
exotic. A partial wave expansion of the $\pi\eta$ system, e.g., will
have components with $L=0,1,2,3,\cdots$ leading to quantum numbers
of the partial waves characterized by $a_0$, $\pi_1$, $a_2$,
$\pi_3$, $\cdots$ where the $\pi_1$ and $\pi_3$ are exotic.
Likewise, $\pi\rho$ in $P$-wave has $1^{-+}$ quantum numbers,
$f_1\pi$ and $b_1\pi$ are $J^{PC}=1^{-+}$ exotic when they are in
$S$-wave. The corresponding isoscalar states are called $\eta$,
$\eta_1$, $\eta_2$, $\cdots$.

Exotic mesons may be hybrid mesons ($q\bar qg$), multiquark states
($q\bar qq\bar q ...$), multimeson states ($M_1~M_2 ...$) or,
possibly, glueballs. Hybrids, tetraquarks (and glueballs) may also
have quantum numbers of ordinary ($q\bar q$) mesons. In this case,
they can mix. In this review, we comment on mesons with exotic
mesons, and on hybrids. The lightest hybrid mesons should have a
mass in the $1.7 - 2.2$ GeV/c$^2$ region even though smaller values
are not ruled out. Tetraquark states should have about the same
mass.

Most experimental information on spin-parity exotic mesons comes
from diffractive or charge exchange scattering of a $\pi^-$ beam off
protons or nuclear targets at fixed beam  momenta (in parentheses),
from E852 at BNL (18 GeV/c; $\eta\pi^-$, $\eta'\pi^-$, $\rho\pi^-$,
$f_1 \pi^-$, $b_1 \pi^-$, $\eta \pi^0$) and VES at Protvino (28, 37
GeV/c; $\eta\pi^-$, $\eta' \pi^-$, $\rho \pi^-$, $f_1\pi^-$,
$b_1\pi^-$, $\eta'\pi^0$). The Crystal Barrel and Obelix
collaborations at LEAR, CERN, have reported evidence for exotic
mesons from $p \bar p$ annihilation at rest ($\eta\pi^{\pm}$, $\eta
\pi^0$, $\rho \pi$, $b_1 \pi$). References to earlier experiments
can be found in\cite{Klempt:2007cp}.

Resonances, hybrids or tetraquark states, and meson-meson molecular
systems can possibly be differentiated. A resonance with
$J^{PC}=1^{-+}$ can decay into $\pi\eta$, $f_1\pi$, $\rho\pi$, and
$b_1\pi$. Of course, the fractions are unknown but there is no
selection rule expected which may suppress one of these decay modes.
If exotic waves originate from diffractive meson-meson scattering,
the $\rho$ may be excited to $b_1$ in $\rho\pi$ scattering but not
$\rho$ to the $\eta$; in $f_1\pi$ scattering, the $f_1$ could be
de-excited to the $\eta$ but not excited to the $b_1$. If
diffractive meson-meson scattering were responsible for the
exotic-wave amplitudes, we might expect different production
characteristics for $\pi\eta$ and $\pi f_1$, and for $\rho\pi$ and
$b_1\pi$.

\subsection{The $\pi_1(1400)$}
\begin{figure}[pt]
\begin{center} \includegraphics[width=0.95\textwidth]{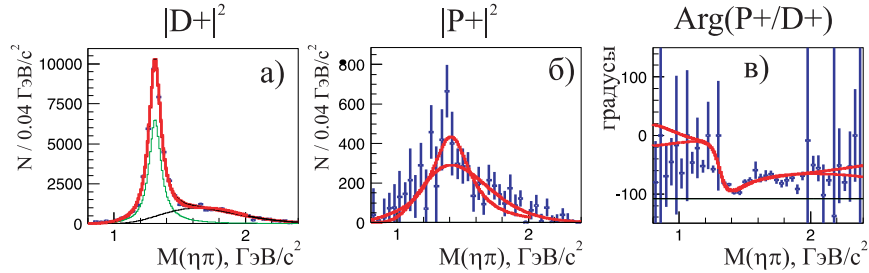}
\caption{ Results of partial-wave analysis of the $\eta \pi^-$
system: a) intensity of $\rm D_+$ wave, ~ b) intensity of $\rm P_+$
wave, c) phase difference $\rm
P_+/D_+$\protect\cite{Dorofeev:2001xu}.}
\label{ex:etpi}
\end{center}
\vspace{-5mm}
\end{figure}

The data in Fig. \ref{ex:etpi} exhibit a dominant $a_2(1320)$ in the
$D_+$ and a clear bump at $M \approx 1.4$\,GeV/c$^2$ in the (exotic)
$P_+$ partial wave. The E-852
collaboration\cite{Thompson:1997bs}\hspace{-1mm}\cite{Chung:1999we}
finds that the data are consistent with a simple ansatz, assuming
contributions from two resonances, one in each partial waves. The
$\rm D_+$ wave returns the parameters of $a_2(1320)$, for the $P_+$
partial wave, mass and width are determined to $M=1370\pm
16~^{+50}_{-30}$~MeV/c$^2$; $\Gamma= 385 \pm 40^{+\ 65}_{-105}$
MeV/c$^2$. A similar fit was used by VES yielding compatible
results\cite{Dorofeev:2001xu}. The VES collaboration tried fits
without resonance but with a phenomenological background amplitude.
The fit gave a significantly worse but not unacceptable $\chi^2$.
Both fits are shown in Fig. \ref{ex:etpi}.

The Indiana group\cite{Szczepaniak:2003vg} used $t$ channel exchange
forces to construct a background amplitude which could mimic
$\pi_1(1400)$. The $\pi\eta$ $P_+$-wave interactions very similar to
$\pi\pi$ $S$-wave interactions were constructed. The latter are
characterized by the $\sigma$ pole; as a consequence, $\pi_1(1400)$
is considered as $\sigma$-type phenomenon in $\pi\eta$ $P_+$-wave
interactions. In the words of the authors
of\cite{Szczepaniak:2003vg}, $\pi_1(1400)$ is `not a QCD bound
state' but rather generated dynamically by meson exchange forces.

Based on $SU(3)$ arguments, a $P$-wave resonance in the $\eta_8 \pi$
channel must belong to a $SU(3)$ decuplet\cite{Chung:2002fz}. The
decuplet-antidecuplet includes also $ K^+ \pi^+$ $P$-wave which
shows practically no phase motion at all\cite{Estabrooks:1977xe}.
Very little phase motion should hence be expected for the $\pi\eta$
$\rm P_+$-wave.

At BNL, the charge exchange reaction $\pi^- p \to \eta \pi^0 n$,
$\eta \to \pi^+ \pi^- \pi^0$ at 18 GeV$/c$ was shown to be
consistent with a resonant hypothesis for the $P_+$ wave, and a mass
of $1257 \pm 20 \pm 25$ MeV$/c^2$, and a width of $354 \pm 64 \pm
60$ MeV$/c^2$ were deduced\cite{Adams:2006sa}. The authors left open
the question if this object should be identified with $\pi_1(1400)$
or if it is a second state in this partial wave. The VES $\eta\pi^0$
spectrum is dominated by the $a_2^0(1320)$ meson; they did not find
evidence for the neutral $\pi_1^0(1600)$.

The Crystal Barrel Collaboration confirmed the existence of the
exotic $\pi\eta$ $\rm P_+$-wave in $ \bar p n \to \pi^- \pi^0
\eta$\cite{Abele:1998gn} and $\bar p p \to 2\pi^0
\eta$\cite{Abele:1999tf}. The Crystal Barrel\cite{Dunnweber:2004vc}
and Obelix\cite{Salvini:2004gz} collaborations found a resonant
contribution of the $J^{PC}=1^{-+}$ wave in $(\rho \pi)$ in $p \bar
p$ annihilation to four pions. However, the $\pi\eta$ $P$-wave is
produced from spin triplet states of the $ N\bar N$ system, the
exotic $\rho\pi$ wave comes from spin singlet states. Hence these
must be different objects, a $\pi_1(1400)$ and a
$\tilde\pi_1(1400)$, plus a neutral $\pi_1(1260)$ if the latter is
another separate resonance. In $p\bar p$ annihilation into
$\pi\pi\eta$, triangle singularities due to final-state rescattering
yield logarithmic divergent amplitudes. The inclusion of
rescattering amplitudes was never attempted; it could possibly
reduce the need for a true pole. In summary, there is evidence for
the existence of a $\pi\eta$ resonance with exotic quantum numbers
but there are severe inconsistencies in the overall picture
associated with its existence.

\subsection{The $\pi_1(1600)$ and $\pi_1(2000)$}
\begin{figure}[pt] \vspace{-12mm}
\includegraphics[width=\textwidth,height=0.7\textwidth]{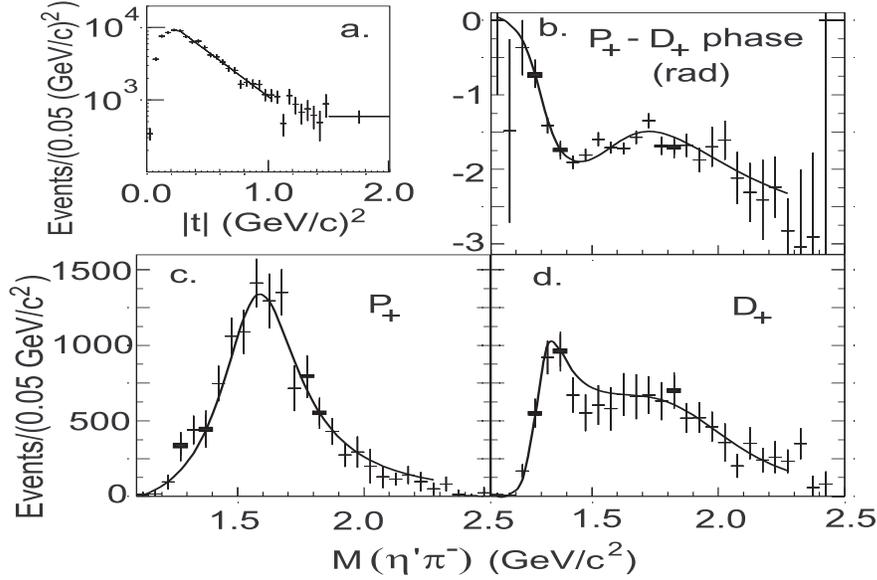}
\caption{\label{ex:et1pi-} BNL data\cite{Chung:1999we}: (a) The
acceptance-corrected $|t|$ distribution fitted with the function
$f(t)=ae^{b|t|}$ (solid line).  (b), (c), (d) results of a
mass-independent PWA and a mass-dependent fit (solid curve) for the
$\rm P_+$ and $\rm D_+$ partial waves and their phase difference.
(b) The $(P_{+}-D_{+})$ phase difference. (c) Intensity of the $P_+$
and (d) of the $D_+$ partial wave. }
\end{figure}

Fig.~\ref{ex:et1pi-} shows the $\eta'\pi^-$ system produced in a
diffractive-like reaction at $p_{\pi_-} =18$\,GeV/c. The data are
from the E-852 collaboration\cite{Ivanov:2001rv}; VES using a beam
at $p_{\pi_-} = 37$\,GeV/c showed similar
distributions\cite{Beladidze:1993km}. The $1^{-+}$ wave exceeds in
intensity the tensor wave and is readily fitted by a Breit-Wigner
resonance at $M \approx 1 600$ MeV/c$^2$ which is listed as
$\pi_1(1600)$ in the Review of Particle Properties. The absence of
$\pi_1^0(1600)$ can be understood by assuming that
\begin{itemize}
\item[-] $\pi_1(1600)$ decouples from $\rho\pi$, or
\item[-] $\pi_1(1600)$ originates from meson-meson diffractive
scattering
\end{itemize}

The wave $J^{PC}=1^{-+}$ in the $\pi^+\pi^-\pi^-$ system was studied
in diffractive-like reactions by the VES
collaboration\cite{Amelin:1995gu}\hspace{-1mm}\cite{Amelin:2005ry}
and by the E-852 collaboration at $p_{\pi} =18$
GeV/c\cite{Adams:1998ff}\hspace{-1mm}\cite{Chung:2002pu} suggesting
the existence of an exotic resonance in $\rho\pi$ which we call
$\tilde\pi_1(1600)$. A new BNL data sample with 10-fold increased
statistics was reported in\cite{Dzierba:2005jg}, yielding negative
evidence for a resonance in the $P_+$ wave. The $\tilde\pi_1(1600)$
must be different from the $\pi_1(1600)$ seen in $\eta'\pi$;
firstly, because of the nearly vanishing coupling of
$\pi_1(1600)\to\rho\pi$ and, secondly, for the different production
modes: the $\pi_1(1600)$ is produced by natural parity exchange, the
$\tilde\pi_1(1600)$ by both, natural and unnatural parity exchange
in about equal portions.

The dominant wave in $f_1 \pi$ is $J^{PC}=1^{-+}$. It is produced
via natural parity exchange; it resembles in production
characteristics the $\eta^{\prime}\pi$ exotic
wave\cite{Amelin:2005ry}. The E-852 collaboration fitted the PWA
intensity distributions and phase differences with a superposition
of Breit-Wigner resonances in all channels.  In the exotic wave, two
resonances are introduced at $M$=(1709$\pm$24$\pm$41),
$\Gamma$=(403$\pm$80$\pm$115)\,MeV/c$^2$ and
$M$=(2001$\pm$30$\pm$92), $\Gamma$=(333$\pm$52$\pm$49)\,MeV/c$^2$
\hspace{-1mm}\cite{Kuhn:2004en}.

Similar observations in $f_1(1285)\pi$ in the
$\omega(\pi^+\pi^-\pi^0) \pi^-\pi^0$ channel studied by
VES\cite{Amelin:2005ry}\hspace{-1mm}\cite{Amelin:1999gk} and
E-852\cite{Lu:2004yn}. Three isobars $\omega \rho,~ b_1 \pi$ and
$\rho_3 \pi$ were considered. The BNL collaboration interprets the
data by resonances, two of them, called $\tilde\pi_1(1600)$ and
$\tilde\pi_1(2000)$ here, are compatible in mass with the findings
from $f_1(1285)\pi$ but are produced via natural and unnatural
parity exchange. The VES data find consistency with a resonance
interpretation but can describe the data without exotic resonances
as well.

\subsection{Conclusions on light-quark exotics}
Partial waves with exotic spin-parity have been observed in several
experiments. The data are consistent with the assumption that the
exotic wave originates from diffractive meson-meson scattering. The
interpretation of the observation as genuine resonances is
controversial.

\begin{figure}[pt]
\begin{center}
\includegraphics[width=1.32\textwidth]{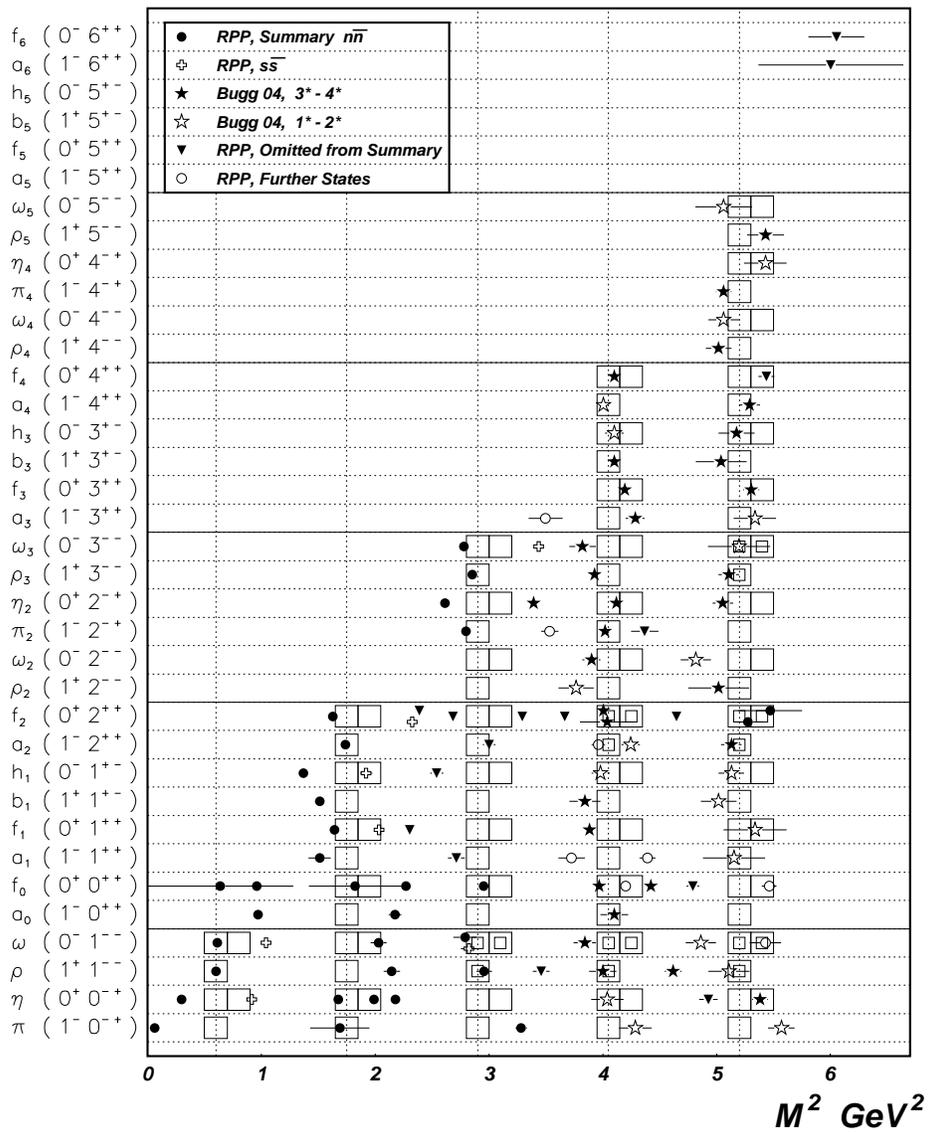}
\end{center}
\vspace{-8mm} \caption{\label{fig:globalview} The pattern of light
quark meson states.}
\end{figure}

\section{Non-exotic hybrid candidates}
\subsection{Light-quark hybrid candidates}
Figure \ref{fig:globalview} shows the light quark mesons with known
quantum numbers $I^G J^{PC}$ as a function of $M^2$. The ordering of
states follows expectations from potential models for $q \bar q$
mesons. The empty boxes indicate the position of states in a
simplified model in which masses of mesons are proportional to
$l+n$, where~$l$~is
\nocite{Barberis:1999be,Barberis:1999wn,Anisovich:2000mv} the
orbital and $n$ the radial quantum number. Mesons of zero isospin
have two nearby boxes for $n \bar n$ and $s \bar s$ states or for
$SU(3)$ singlet and octet states. Some boxes are doubled because two
different states, with $J=l+1,n$ and $J=l-1,n+2$, are expected.
Nearly all observed mesons are compatible with a $q \bar q$
assignment,  with two remarkable exceptions, $\pi_2(1870)$ and
$\eta_2(1870)$\cite{Adomeit:1996nr}$^{-}\hspace{-1mm}$\cite{Anisovich:2001hj}.
These two states are meaningful hybrid candidates. When scrutinizing
these observation, we notice that comparatively narrow hybrids are
predicted for the $J^{PC}=2^{-+}$ wave which has important $S$-wave
thresholds, $f_2(1270)\pi$ and $a_2(1320)\pi$. Narrow hybrids are
also predicted for the $J^{PC}=1^{++}$ wave. In this wave, there are
no important $S$-wave thresholds, and no hybrid candidates, neither.
Certainly, a good understanding of the threshold dynamics is
required. High statistics data in several final states are mandatory
to resolve this issue.

\subsection{Is there restoration of
chiral symmetry?} There is a degeneracy of the masses with positive
and negative parities which has been interpreted as evidence for
restoration of chiral symmetry in highly excited
mesons\cite{Glozman:2002cp}$^{-}\hspace{-1mm}$\cite{Glozman:2004gk}.
A new QCD scale $\Lambda_{CSR}=2.5$\,GeV/c$^2$ is suggested at which
chiral symmetry is
restored\cite{Swanson:2003ec}\hspace{-1mm}\cite{Afonin:2006vi}.

The $l,n$ degeneracy follows also from a model based on the dual
superconductor mechanism of confinement\cite{Baker:2002km} and from
a model guided by the correspondence of the dynamics of quarks in
QCD and of strings in a five-dimensional Anti-de-Sitter
space\cite{Karch:2006pv}. Both approaches suggest
$$ M_n^2(l) = 2\pi\sigma\left(l+n+\frac{1}{2}\right)\,. $$ which shows
that the squared masses are linear in $l$ and $n$, and degenerate in
$n+l$. The string model and the conjectured restoration of chiral
symmetry thus both lead to a $n+l$ degeneracy of excited states. The
two models make however different predictions for `stretched'
states, for states with $J=l+s$. The string model predicts no parity
partners for $a_2(1320)$--$f_2(1270)$,
$\rho_3(1690)$--$\omega_3(1670)$, $a_4(2040)$--$f_4(2050)$,
$\rho_5(2350)$--$\omega_5$, $a_6(2450)$--$f_6(2510)$ while their
existence should be expected if chiral symmetry restoration is at
work. Experimentally, there are no chiral partner for any of these
10 states. Hence, at the first glance, data do not support the
hypothesis of chiral symmetry restoration.

\subsection{J/$\psi$ excitations}
With the discovery of the $h_c(1P)$ and $\eta_c(2S)$ resonances, an
important milestone was reached: all charmonium states predicted by
quark models below the $D\bar D$ threshold have been found, and no
extra state. Above the $D\bar D$ threshold, several surprisingly
narrow states were found called $X(3872)$, $X(3940)$, $Y(3940)$, and
$Z(3930)$. In spite of some anomalous properties, these states can
be assigned to $\chi_1(2P)$, $\eta_c(3S)$, $\chi_0(2P)$, and
$\chi_2(2P)$. Reasons for this assignment are discussed
in\cite{Klempt:2007cp}. A particularly demanding state is the
$Y(4260)$ which is discussed next.

\subsection{The $Y(4260)$}
The $Y(4260)$ was discovered by the BaBaR collaboration as an
enhancement in the $\pi\pi J/\psi$ subsystem in the initial state
radiation (ISR), in $e^+e^- \to \gamma_{\rm ISR}+$ $
J/\psi\pi\pi$\cite{Aubert:2005rm}. Its mass was determined to $4259
\pm 8 \pm 4$ MeV/c$^2$, the width to $88 \pm 23 \pm 5$ MeV/c$^2$,
the spin-parity to $J^{PC}=1^{--}$. The $Y(4260)$ resonance was
searched for in the inclusive $e^+e^-$ annihilation cross
section\cite{Mo:2006ss}. In the $\sqrt s=4.20-4.35$\,GeV/c$^2$
region, the cross section exhibits a dip-bump-dip structure which
makes it difficult to extract a reliable estimate for a possible
$Y(4260)$ contribution. The apparent absence of $Y(4260)$ in this
reaction has stimulated the interpretation that it could be a
hybrid\cite{Zhu:2005hp}$^{-}\hspace{-1mm}$\cite{Kou:2005gt} or a
tetraquark resonance\cite{Maiani:2005pe}. The upper limit of
$Y(4260)$ in the inclusive $e^+e^-$ annihilation cross section
depends however on the flexibility of the fit. If constructive and
destructive interferences are allowed, the upper limit for $Y(4260)$
is less stringent and a scenario as suggested in Table
\ref{psiprimes} is not excluded.

At this conference, W.S.~Hou reported observation of $e^+e^- \to
\Upsilon(1S)\pi^+\pi^-$, $\Upsilon(2S)\pi^+ \pi^-$, and
$\Upsilon(3S)\pi^+\pi^-$  at $\sqrt{s}~10.87$ GeV, near the peak of
the $\Upsilon(10860)$. If these signals originate from the
$\Upsilon(10860)$ resonance, the corresponding partial widths are
much larger than expected and would suggest that $\Upsilon(10860)$
-- and $Y(4260)$ as well -- be a hybrid.

\begin{table}[pt]
\caption{\label{psiprimes}Charmonium states with $J^{PC}=1^{--}$ in
our interpretation. The partial widths are given in keV/c$^2$, the
masses in MeV/c$^2$.\vspace{-2mm}}
\begin{center}
\begin{scriptsize}\renewcommand{\arraystretch}{1.7}
\begin{tabular}{cccccccc}
\hline\hline
J/$\psi$\hspace{-5mm}&\hspace{-5mm}$\psi(3686)$\hspace{-3mm}&\hspace{-3mm}$\psi(3770)$\hspace{-3mm}&\hspace{-3mm}$\psi(4040)$\hspace{-3mm}&\hspace{-3mm}$\psi(4160)$\hspace{-3mm}&\hspace{-3mm}$Y(4260)$\hspace{-3mm}&\hspace{-3mm}$\psi(4415)$\\
\hspace{-3mm}&\hspace{-3mm}\quad$2S$\ \quad\hspace{-3mm}&\hspace{-3mm}\quad$1D$\quad\hspace{-3mm}&\hspace{-3mm}\quad$3S$\quad\hspace{-3mm}&\hspace{-3mm}\quad$2D$\quad\hspace{-3mm}&\hspace{-3mm}\quad$4S$\quad\hspace{-3mm}&\hspace{-3mm}\quad$5S$ \\
\hline
$\Gamma_{e^+e^-}$\hspace{-5mm}&\hspace{-5mm}$2.48\pm0.06$\hspace{-3mm}&\hspace{-3mm}$0.242^{+0.027}_{-0.024}$\hspace{-3mm}&\hspace{-3mm}$0.86\pm0.07$
\hspace{-3mm}&\hspace{-3mm}$0.83\pm0.07$\hspace{-3mm}&\hspace{-3mm} {\it 0.72} \hspace{-3mm}&\hspace{-3mm}$0.58\pm0.07$\\
$\Gamma_{\rm
J/\psi\pi^+\pi^-}$\hspace{-5mm}&\hspace{-5mm}$107\pm5$\hspace{-3mm}&\hspace{-3mm}$44\pm8$\hspace{-3mm}&\hspace{-3mm}
$<360$ \hspace{-3mm}&\hspace{-3mm}$<330$\hspace{-3mm}&\hspace{-3mm} $670\pm240$ \hspace{-3mm}&\hspace{-3mm}-\\
$M_{\psi(nS)}-M_{J/\psi}$  \hspace{-5mm}&\hspace{-5mm} \quad589\
\quad\hspace{-3mm}&\hspace{-3mm}\quad674\
\quad\hspace{-3mm}&\hspace{-3mm}\quad943\quad\hspace{-3mm}&\hspace{-3mm}\quad1056\
\quad\hspace{-3mm}&\hspace{-3mm}\quad1163\quad\hspace{-3mm}&\hspace{-3mm}\quad1318\\
$M_{\Upsilon(nS)}-M_{\Upsilon}$ \hspace{-5mm}&\hspace{-5mm}
\quad563\ \quad\hspace{-3mm}&\hspace{-3mm}\hspace{-3mm}&\hspace{-3mm}\quad895\quad\hspace{-3mm}&\hspace{-3mm}\hspace{-3mm}&\hspace{-3mm}\quad1119\quad\hspace{-3mm}&\hspace{-3mm}     \\
\hline\hline
\renewcommand{\arraystretch}{1.0}
\end{tabular}
\end{scriptsize}
\end{center}
\vspace{-10mm}
\end{table}

\subsection{Is there a $\psi(2S)\pi^+$}
A narrow $\psi'\pi^{\pm}$ resonance was observed by the Belle
collaboration in $B$ decays to $K \pi^+\psi'$, with a statistical
evidence exceeding 7$\sigma$\cite{:2007wg}. The resonance, called
$Z^+(4430)$, has $4433\pm 4\pm 1$\,MeV/c$^2$ mass and a width of
$\Gamma = 44^{+17}_{-13}$$^{+30}_{-11}$\,MeV/c$^2$. It is the first
charged resonance with hidden charm; evidently, it can not belong to
the charmonium family. It was interpreted as tetraquark radial
excitation\cite{Maiani:2007wz}. Rosner noticed that the $Z(4430)$
mass is at the $D^*\bar{D}_1(2420)$ threshold and proposed that the
state is formed via the weak $b\to c\bar cs$ transition, creation of
a light-quark pair, and rescattering of the final-state
hadrons\cite{Rosner:2007mu}. Hence at present, there not yet the
need for an interpretation beyond the standard quark model using
$q\bar q$ only.

\section{Conclusions and outlook}
In the view presented here, there is not yet a convincing answer to
the question if hybrid mesons exist. When data are analyzed assuming
the existence of hybrids, evidence is observed in several places. If
this conjecture is examined with scrutiny, the evidence for hybrids
fades away. There are, however, specific predictions for the outcome
of future experiments. If exotic partial waves are due to
diffractive meson-meson scattering, the $\pi_1$ partial wave should
not be produced in the charge exchange reaction $\pi^- p\to
n\pi_1^0(1400)$. The $\pi_1^0(1400)$ observed in\cite{Adams:2006sa}
is in conflict with this conjecture, but in conflict with VES data,
too. Likewise, there should be no production of $\pi_1^0(1600)$ or
$\pi_1^0(2000)$. In central production, a large contribution to the
cross section will come from Regge-Pomeron fusion which should be a
good place to search for hybrids. With two detected protons, no
charged Reggeon is exchanged (with Reggeon exchange = Regge or
Pomeron exchange); diffractive meson-meson scattering leads to
neutral final states and no hybrids with exotic quantum numbers
should be found. At Jlab, the initial state $\gamma p$ is charged,
and partial waves with exotic quantum numbers due to diffractive
meson-meson scattering should be observed in their charged state
only.

The new BELLE results on $\Upsilon(10860)$ decays reported by
W.S.~Hou are very suggestive. If the signals are due to an extremely
large $\Upsilon(10860)\to \Upsilon(nS)\pi^+\pi^-$ decay mode,
hybrids with hidden beauty seem to be a natural consequence.
Similarly, the $Y(4260)$ might be of hybrid nature as well. Hence
there is room left; Panda at GSI (or, earlier, BELLE) will have to
give us the final answer. The existence or not of glueballs -- which
were not discussed here -- is a question which should find its
answer from BESIII.

\end{document}